%% file: main.tex
\documentclass[11pt,a4paper]{article}

\usepackage[utf8]{inputenc}
\usepackage[T1]{fontenc}
\usepackage[a4paper,margin=1.1in]{geometry}
\usepackage{newtxtext} 
\usepackage{amsmath} 
\usepackage{mathtools}
\usepackage{amsfonts}
\usepackage{mathrsfs}
\usepackage{dutchcal}
\usepackage{tikz}
\usetikzlibrary{shapes.misc, positioning}
\usepackage{ntheorem} 
\usepackage{microtype} 
\usepackage{xcolor}
\usepackage{url}
\usepackage{booktabs}
\usepackage{nicefrac}
\usepackage{hyperref}
\usepackage{todonotes}

\input{header}
\input{macros}
\input{header-gpt-experiments}
\input{macros-gpt-experiments}

\newenvironment{ack}{
  \section*{Acknowledgements}
}{}

\title{The KG-ER Conceptual Schema Language}

\author{
Enrico Franconi \\
Free University of Bozen-Bolzano, Italy\\
\texttt{franconi@inf.unibz.it}
\and
Beno\^{\i}t Groz \\
Universit\'{e} Paris-Saclay, CNRS, LISN, France\\
\texttt{groz@lisn.fr}
\and
Jan Hidders \\
Birkbeck, University of London, UK \\
\texttt{j.hidders@bbk.ac.uk}
\and
Nina Pardal \\
University of Huddersfield, UK \\
\texttt{n.pardal@hud.ac.uk}
\and
S{\l}awek Staworko\thanks{Corresponding author.} \\
Relational AI, Berkeley, CA, USA \\
\texttt{slawek.staworko@relational.ai}
\and
Jan Van den Bussche \\
Hasselt University, Hasselt, Belgium\\
\texttt{jan.vandenbussche@uhasselt.be}
\and
Piotr Wieczorek \\
University of Wrocław, Poland \\
\texttt{piotr.wieczorek@cs.uni.wroc.pl}
}

\date{}

\begin{document}

\maketitle

\input{intro}
\input{kg-er}
\input{data-model}
\input{semantics}
\input{disjointness}

\input{future}
\input{acknowledgements}

\small
\bibliographystyle{plain}
\bibliography{bibliography}

\newpage
\appendix
\input{appendix}

\input{appendix-translation}

\end{document}

%% file: header.tex
\usepackage{newtxtext}
\usepackage{amsmath} 
\usepackage{mathrsfs}
\usepackage[utf8]{inputenc}
\usepackage{dutchcal}
\usepackage{tikz}
\usepackage{ntheorem}
\usepackage{mathtools}
\usetikzlibrary{shapes.misc}
\usetikzlibrary{positioning}
\usepackage{todonotes}

\usepackage{listings}
\lstdefinelanguage{PGSchema}
{
  morekeywords={
    NODE,
    EDGE,
    TYPE,
    IMPORTS,
    OPTIONAL,
    DATE,
    INT,
    STRING,
    BOOL,
    ARRAY,
    ENUM,
    DOUBLE,
    INT32,
    DATETIME,
    STRICT,
    LOOSE,
    OPEN,
    CLOSED,
    ABSTRACT,
    CREATE,
    GRAPH,
    FOR,
    WITHIN,
    EXCLUSIVE,
    MANDATORY,
    SINGLETON,
    FLOAT,
    IDENTIFIER,
    MATCH,
    WHERE,
    NOT,
    OR,
    AND,
    EXISTS,
    RETURN,
    IN,
    IS,
    NULL,
    hasKey,
    FunctionalProperty,
    context,
    test,
    unique,
    selector,
    field,
    allInstances,
    isUnique,
    Tuple,
    class,
    extent,
    relationship,
    inverse,
    attribute,
    COUNT,
    OF
  },
  sensitive=true, 
  morecomment=[l]{//}, 
  morecomment=[s]{/*}{*/}, 
  morestring=[b]" 
}

\usepackage{color}
\definecolor{eclipseBlue}{RGB}{42,0.0,255}
\definecolor{eclipseGreen}{RGB}{63,127,95}
\definecolor{eclipsePurple}{RGB}{127,0,85}
 

%% file: macros.tex
\newcommand{\Ent}{\mathcal{Ent}}
\newcommand{\Rel}{\mathcal{Rel}}
\newcommand{\Attr}{\mathcal{Attr}}
\newcommand{\Rol}{\mathcal{Rol}}
\newcommand{\Entity}{\textsc{Entity}}
\newcommand{\Relationship}{\textsc{Relationship}}
\newcommand{\Attribute}{\textsc{Attribute}}
\newcommand{\Role}{\textsc{Role}}
\newcommand{\Mandatory}{\textsc{Mandatory}}
\newcommand{\Single}{\textsc{Single}}
\newcommand{\Key}{\textsc{Key}}
\newcommand{\Id}{\textsc{Identity}}
\newcommand{\Isa}{\textsc{Isa}}
\newcommand{\Disjoint}{\textsc{Disjoint}}
\newcommand{\Cover}{\textsc{Cover}}

\newcommand{\V}{\mathcal{V}}

\newcommand{\D}{\mathcal{D}}
\newcommand{\ent}{\mathrm{ent}}
\newcommand{\rel}{\mathrm{rel}}
\newcommand{\fin}{\mathrm{fin}}
\renewcommand{\L}{\mathcal{L}}
\newcommand{\G}{\mathcal{G}}
\newcommand{\sem}[1]{[\![#1]\!]}
\newcommand{\disjoint}{{\mathord{\circ\kern-0.5pt\circ}}}
\newcommand{\overlapping}{{\mathord{\circ\kern-2.5pt\circ}}}
\newcommand{\WF}[1]{$\mathbf{WF}_{#1}$}

\def\qed {{                
   \parfillskip=0pt        
   \widowpenalty=10000     
   \displaywidowpenalty=10000  
   \finalhyphendemerits=0  
   \leavevmode             
   \unskip                 
   \nobreak                
   \hfil                   
   \penalty50              
   \hskip.2em              
   \null                   
   \hfill                  
   $\square$
   \par}}                  

\tikzset{
    entity/.style={
        rectangle,
        draw,
        minimum width=2cm,
        minimum height=0.7cm,
        thick
    },
    attribute/.style={
        rounded rectangle,
        draw,
        minimum width=1.5cm,
        minimum height=0.5cm,
        thick
    },
    relationship/.style={
        chamfered rectangle,
        chamfered rectangle angle=25,
        draw,
        minimum width=1.75cm,
        minimum height=0.65cm,
        thick,
        chamfered rectangle xsep=2cm
    },
    dot/.style={
        circle,
        draw,
        minimum size=0.15cm,
        inner sep=0cm,
    }
}

\colorlet{Orange}{orange!80!black}
\colorlet{Blue}{blue!60!black}
\colorlet{Green}{green!60!black}
\colorlet{Red}{red!60!black}

\theoremstyle{plain} 
\theoremheaderfont{\bfseries\upshape} 
\theoremseparator{.} 
\theorembodyfont{\normalfont}
\newtheorem{example}{Example}[section]

%% file: header-gpt-experiments.tex
\usepackage{listings}
\usepackage[skins,breakable,listings]{tcolorbox}
\usepackage{xcolor}

\lstdefinestyle{sqlstyle}{
  language=SQL,
  basicstyle=\ttfamily\small,
  keywordstyle=\color{blue}\bfseries,
  commentstyle=\color{gray},
  stringstyle=\color{red},
  showstringspaces=false,
  frame=single,
  breaklines=true
}


%% file: macros-gpt-experiments.tex
\newtcblisting{sql}{
listing engine=listings,
colframe=white,
boxrule=.2mm,
listing only,
listing options={
  language=SQL,
  basicstyle=\ttfamily\small,
  keywordstyle=\color{blue}\bfseries,
  commentstyle=\color{gray},
  stringstyle=\color{red},
  tabsize=2,
  showstringspaces=false
},
left=.1mm,
top=.1mm,
bottom=.1mm
}

\newtcblisting{sqlbreakable}{%
  listing engine=listings,
  listing only,
  breakable,
  colframe=white,
  boxrule=.2mm,
  left=.1mm,
  top=.1mm,
  bottom=.1mm,
  listing options={%
    language=SQL,
    tabsize=2,
    basicstyle=\footnotesize\ttfamily,
    breaklines=true
  }
}

\newtcblisting{gptexplanation}{
listing engine=listings,
colframe=white,
coltext=gray,
boxrule=.2mm,
listing only,
listing options={
  language={}, 
  basicstyle=\ttfamily\scriptsize,
  tabsize=2,
  showstringspaces=false
},
left=.1mm,
top=.1mm,
bottom=.1mm
}

\newtcblisting{markdown}{
listing engine=listings,
colframe=white,
boxrule=.2mm,
listing only,
listing options={
  language={}, 
  basicstyle=\ttfamily\small,
  tabsize=2,
  showstringspaces=false
},
left=.1mm,
top=.1mm,
bottom=.1mm
}

\newtcblisting{conceptual}{
listing engine=listings,
colframe=white,
boxrule=0mm,
listing only,
listing options={
  language={}, 
  basicstyle=\ttfamily\small,
  tabsize=2,
  showstringspaces=false
},
left=0mm,
top=0mm,
bottom=.1mm
}

\newtcolorbox{chatbox}{
colback=white,
enhanced,
breakable,
drop fuzzy shadow=black!50,
interior style={white},
boxrule=.5pt,toprule=.5pt,bottomrule=.5pt, left=0pt,right=0pt,top=1pt,bottom=1pt,
colframe=black
}

\newtcolorbox{promptbox}[1][]{
colbacktitle=gray!50,coltitle=black,colback=white,
enhanced,
drop fuzzy shadow=black!50,
interior style={white},
boxrule=.5pt,titlerule=0pt,toprule=.5pt,bottomrule=.5pt, left=0pt,right=0pt,top=1pt,bottom=1pt,
colframe=black,title={\bfseries Prompt:},#1
}

\newtcolorbox{gptbox}[1][]{
colbacktitle=gray!50,coltitle=black,colback=white,
sharpish corners=all,
enhanced,
drop fuzzy shadow=black!50,
interior style={white},
boxrule=.5pt,titlerule=0pt,toprule=.5pt,bottomrule=.5pt, left=0pt,right=0pt,top=1pt,bottom=1pt,
colframe=black,title={\bfseries GPT-4o:},#1
}

%% file: intro.tex
\begin{abstract}
  We propose \emph{KG-ER}, a conceptual schema language for knowledge graphs that describes
  the structure of knowledge graphs independently of their representation (relational
  databases, property graphs, RDF) while helping to capture the semantics of the information
  stored in a knowledge graph.
\end{abstract}

\section{Introduction}
\label{sec:intro}

Knowledge graphs (KGs) have become central to many AI
applications~\cite{thalhammer2022knowledge}, benefiting many AI-based tasks, including
NLP and reasoning~\cite{Mirasdar2025KnowledgeGraphs}, data
integration~\cite{Liang2022SurveyKGR,farah2025knowledge}, and semantic
search~\cite{verma2024beyond}. KGs organize information as graphs, with nodes as entity
instances and edges as relationship instances~\cite{hogan2021knowledge,chaudhri2022knowledge}. Systems like
property graphs~\cite{gheerbrant2024gqlsqlpgqtheoreticalmodels} and RDF~\cite{W3C_RDF} are
built on this principle, and even relational databases can be viewed as
KGs~\cite{ndefo:franconi:ijsc-20}: they too store information about objects and their
relationships, although the information may be structured in a more arbitrary manner.

To use KGs effectively, it is essential to describe their structure and semantics. Database
schemas define expected data structures, e.g., specifying that a \textit{Book} is
\textit{written} by an \textit{Author} and has a unique \textit{ISBN}.  Conceptual modeling
languages—such as ER~\cite{Chen1976,di_battista_deductive_1993}, ORM2~\cite{franconi2020orm,Halpin-book:2024}, and UML~\cite{Fowler2003,berardi_reasoning_2005}—provide intuitive, high-level
interpretations of these structures. The line between schemas and conceptual models is often
blurred, as conceptual modeling languages frequently serve to define schemas with
domain-relevant, intuitive names for object classes (entities) and relationships.

However, supported schema features vary across systems~\cite{10.1007/978-3-319-25264-3_45} and can even be tied to the
underlying data model. Consequently, one cannot assume that database schema is sufficiently
expressive to fully capture the structure and the semantics of the underlying knowledge
graph.

To address this deficiency we propose \emph{KG-ER} (Section~\ref{sec:kg-er}), a conceptual
schema language that serves both as a schema capturing the structure of knowledge graphs and
a modeling language to convey their semantics.  KG-ER has been designed with the aim to provide 
modeling features particularly suited for knowledge graphs, and in particular, it supports:
\begin{itemize}
    \itemsep0pt
    \item Entity types, with support for fine-grained inheritance (disjointness and
    totality) and expressive key constraints based on tree-patterns (acyclic conjunctive queries).
    \item Relationship types, of arbitrary arity, with support for multi-edge relationships,
    key constraints (known as determinants in TigerGraph~\cite{lee2023graph}), and participation
    constraints.
    \item Attributes, for entities and relationships, with support for multi-valued
    attributes, as well as mandatory and single-valued attributes.
\end{itemize}

We intentionally omit less commonly embraced concepts, such as cardinality constraints or
inheritance between relationships; previous research~\cite{DBLP:journals/pvldb/BonifatiDFHHMMS22, DBLP:journals/sigmod/BonifatiD0H0MMS23, DBLP:conf/icdt/GrozLSW22} has 
revealed that these features are indeed rarely used in practice. We also pragmatically limit the scope of 
certain concepts, such as general disjointness of entities, whose interpretation in the context of knowledge
graphs leads to a complex and nuanced debate, as presented in Section~\ref{sec:disjointness}. 


If we compare KG-ER to major existing knowledge graph data models, we can observe the following. Compared to ER and EER data models it has extra features such multi-edge semantics for relationships, a more powerful notion of keys, more restricted participation constraints and it does not allow relationship hierarchies.
Compared to PG-Schema it has simpler key constraints, lacks cardinality constraints and union types, although the latter can be simulated. 
Compared to SHACL and ShEx (which are in some ways similar and in some ways different from SHACL \cite{ahmetaj2025foundationsshapelanguage}) it lacks features such as complex constraints based on regular path queries and constraints that involve nested quantifiers and boolean operators, but it does add composite keys.
We argue that KG-ER because of its choice of features is eminently suitable for discussing and designing knowledge graphs stored in different logical data models such as RDF, Property Graphs and relational databases.

%

As evidence of the usefulness and expressiveness of KG-ER, we note that the schema of the LDBC-SNB
benchmark~\cite{angles2020ldbc} can be captured using KG-ER. Naturally, KG-ER
can be extended with support for additional features, if needed. 

We also provide, in Sections~\ref{sec:data-model} and~\ref{sec:semantics} a rigorous formal
semantics for KG-ER, which can be used to map KG-ER to existing schemas for knowledge graph
representations, such as property graphs schemas~\cite{PG-schema}, schemas for RDF such as ShEx and SHACL~\cite{Pareti2022, DBLP:conf/icdt/StaworkoBGHPS15},
and relational schemas in various normal forms. Because KG-ER is a conceptual schema language, it is 
conceivable to use it for constructing mappings and transformations between knowledge graphs stored using different representations e.g., an RDF with a SHACL to a Property Graph with a PG-Schema.

There are many benefits of using KG-ER in the context of AI practice and research alike.
For AI practitioners, the KG-ER, which consists of a set of simple statements, can be easily
fed into a specific AI model. 
In the full version of the paper~\cite{franconi2025kgerconceptualschemalanguage} we 
we illustrate precisely this benefit by
verbalizing KG-ER in helping LLMs in solving a number of common database tasks (text-to-query, query optimization, and schema normalization).
For AI theoreticians, the precise logical
formalization of KG-ER provides a yardstick of the expressive power for a given AI model to
possess, should it need to operate on the structural and semantic information about the
knowledge graph.

%% file: kg-er.tex
\section{KG-ER Schemas}
\label{sec:kg-er}
Throughout this paper we assume pairwise disjoint sets of names of entities $\Ent$,
relationships $\Rel$, attributes $\Attr$, and roles $\Rol$.

A \emph{KG-ER schema} (or simply \emph{schema}) is a set $S$ of simple statements that we
divide into two parts: the shape graph and the constraints. For every kind of statement we
provide an intuitive natural language verbalization, and we provide precise formal
semantics in Sections~\ref{sec:data-model} and~\ref{sec:semantics}. We consider
only \emph{well-formed} schemas which satisfy certain structural conditions, stated
alongside the introduced statements (using the $\mathbf{WF}$ indicator).

\subsection{Shape Graphs}
\label{sec:shape-graphs}
A shape graph of a schema consists of simple structural elements that describe the basic
topology of knowledge graphs.

The shape graph part of a schema $S$ is a collection of statements of one of the
following forms (with $E\in\Ent$, $R\in\Rel$, $X\in\Ent\cup\Rel$, $A\in\Attr$, and
$B\in\Rol$):
\begin{description}
  \itemsep0pt
  \item $\Entity(E)$: \textit{$E$ is an entity};
  \item $\Relationship(R)$: \textit{$R$ is a relationship};
  \item $\Attribute(X, A)$: \textit{$A$ is an attribute of the entity/relationship $X$};
  \item $\Role(R, B, E)$: \textit{$E$ participates in $R$ in the role $B$}.
\end{description}
\WF{1}: \emph{All attribute and role declarations of a shape graph only use entities
and relationships declared in the shape graph.} Formally, for every $\Attribute(X, A)$ in the
shape graph, it also contains $\Entity(X)$ or $\Relationship(X)$, and for every $\Role(R, B,
E)$ in the shape graph, it also contains $\Relationship(R)$ and $\Entity(E)$.

\begin{example}
\label{ex:shape-graph}
As an example consider the following set of statements that describe a simple shape graph
with persons studying at universities.
\begin{normalfont}
\begin{align*}
  &\Entity(\textit{University}),& &\Attribute(\textit{University}, \textit{name}),\\
  &\Entity(\textit{Person}),& &\Attribute(\textit{Person}, \textit{fname}),\\
  &\Attribute(\textit{Person}, \textit{lname}),& &\Attribute(\textit{Person}, \textit{email}),\\
  &\Relationship(\textit{studies}),& &\Attribute(\textit{studies}, \textit{year}),\\
  &\Role(\textit{studies}, \textit{uni}, \textit{University}),& &
  \Role(\textit{studies}, \textit{student}, \textit{Person}).\tag*{$\square$}
\end{align*}
\end{normalfont}
\end{example}

For simplicity, we assume global uniqueness of attribute and role names in the schema i.e., no attribute name
appears more than once (with different entities or relationships) and similarly no role name
appears more than once (with different relationships or connected entities). Note that attribute names and role names are also assume not to overlap.  The shape graph
in Example~\ref{ex:shape-graph} satisfies this assumption but it would violate it if it also contained
$\Attribute(\textit{University}, \textit{email})$.

\paragraph{Relational vocabulary}

The shape graph of a schema $S$ defines the finite vocabulary $\L_S$ that consists of the
set of entity names $\Ent_S$, relationship names $\Rel_S$, attribute names $\Attr_S$, and
role names $\Rol_S$ used in $S$.  For simplicity of notation, we assume a fixed shape
graph, and in the remaining, we consistently use $E_i$, with $1 \leq i \leq n$, to range over $\Ent_S$,
$R$ to range over $\Rel_S$, $X$ to range over $\Ent_S\cup\Rel_S$, $A$ to range over
$\Attr_S$, and $B$ to range over $\Rol_S$.

\subsection{Graphical Representation and Pattens}

\paragraph{Graphical Representation}
A shape graph is represented diagrammatically by using different node shapes and
labels for both nodes and edges. We use rectangular nodes for entities, oval nodes for
attributes, and hexagonal nodes for relationships. The node label is the name of the
respective element. Additionally, role names are used as labels for the edges that connect
relationships to entities.

  \newbox{\shapegraph}
  \savebox{\shapegraph}{
  \begin{tikzpicture}
    \node[entity] (University) at (0.1,0) {\it University};
    \node[attribute] (name) at (-0.25,1) {\it name};

    \node[relationship] (studies) at (3,0) {\it studies};
    \node[attribute] (year) at (3.25,-1) {\it year};

    \node[entity] (Person) at (6.25,0) {\it Person};
    \node[attribute] (fname) at (7.75,1) {\it fname};
    \node[attribute] (lname) at (6.25,1) {\it lname};
    \node[attribute] (email) at (4.75,1) {\it email};

    \node[relationship] (follows) at (6.25,-1.75) {\it follows};
    \node[attribute] (since) at (8,-2) {\it since};

    \node[relationship] (wrote) at (9.5,0) {\it wrote};

    \node[entity] (Message) at (12.5,0) {\it Message};
    \node[attribute] (number) at (11.5,1) {\it number};
    \node[attribute] (date) at (13,1) {\it date};
    \node[attribute] (text) at (13,-1) {\it text};

    \draw (University) edge[-, thick] (name);

    \draw (year) edge[-, thick] (studies);

    \draw (University) edge[-, thick] node[above] {\it uni} (studies);
    \draw (studies) edge[-, thick] node[above] {\it student} (Person);

    \draw (fname) edge[-, thick] (Person);
    \draw (lname) edge[-, thick] (Person);
    \draw (email) edge[-, thick] (Person);

    \draw (Person) edge[-, bend left, thick] node[right] {\it follower} (follows);
    \draw (follows) edge[-, bend left, thick] node[left] {\it followee} (Person);
    \draw (since) edge[-, thick] (follows);

    \draw (Person) edge[-, thick] node[above] {\it author} (wrote);
    \draw (wrote) edge[-, thick] node[above] {\it msg} (Message);

    \draw (number) edge[-, thick] (Message);
    \draw (date) edge[-, thick] (Message);
    \draw (text) edge[-, thick] (Message);

  \end{tikzpicture}
  }

\begin{example}
\label{ex:shape-graph-graphical}
Graphical representations are conveniently
\begin{figure}[htb]
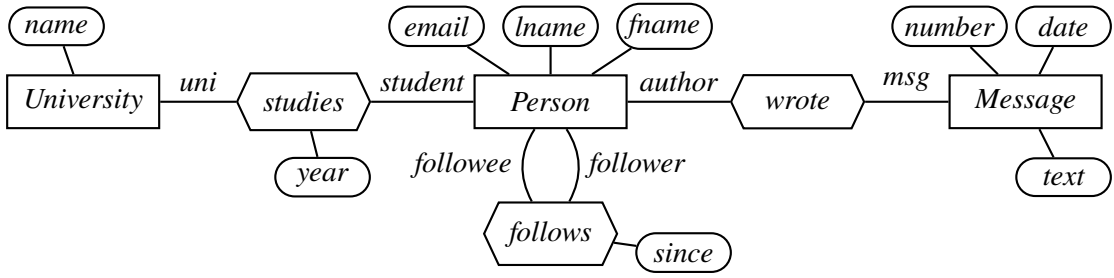

  \small
  \centering
  \usebox{\shapegraph} 
  \caption{Shape graph inspired by the LDBC-SNB schema. \label{fig:shape-graph}}
\end{figure}
more compact as illustrated in Figure~\ref{fig:shape-graph}, which extends the shape graph
in Example~\ref{ex:shape-graph}. \qed
\end{example}

\paragraph{Patterns}
We use \emph{tree patterns}, a simple yet powerful querying metaphor, to specify the key
information that constitutes the identifying information of entity and relationship
instances. Formally, a \emph{pattern} over a schema $S$ is a term $p$ over the set of
attribute names $\Attr_S$ and role names $\Rol_S$, with the attribute names appearing only
in the leaf nodes. The \emph{arity} of a pattern is the number of its leaf nodes.

We use patterns to query knowledge graphs whose structure conforms to the shape graph, and
so we only use valid patterns that navigate through the shape graph following the existing
edges labeled with the given role names and may terminate at existing attributes. Patterns
have a natural meaning, which is best illustrated graphically as shown on the example below.
\begin{example}
\label{ex:patterns}
We illustrate patterns with graphical examples in Figure~\ref{fig:patterns}, which contains a
fragment of the shape graph from Example~\ref{ex:shape-graph-graphical}.
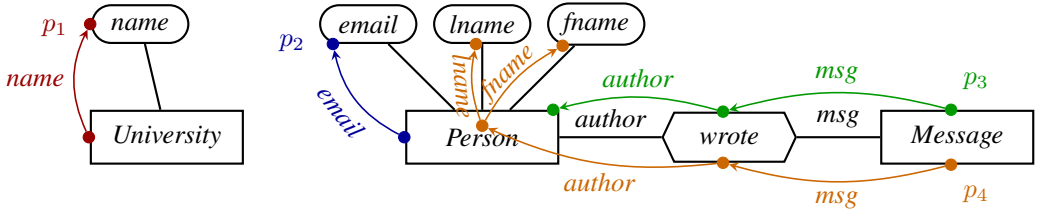
\begin{figure}[htb]
  \small
  \centering
  \begin{tikzpicture}[>=stealth]
    \node[entity] (University) at (2.1,0) {\it University};
    \node[attribute] (name) at (1.75,1.5) {\it name};

    \node[entity] (Person) at (6.25,0) {\it Person};
    \node[attribute] (fname) at (7.75,1.5) {\it fname};
    \node[attribute] (lname) at (6.25,1.5) {\it lname};
    \node[attribute] (email) at (4.75,1.5) {\it email};

    \node[relationship] (wrote) at (9.5,0) {\it wrote};
    \node[entity] (Message) at (12.5,0) {\it Message};
    \draw (University) edge[-, thick] (name);

    \draw (fname) edge[-, thick] (Person);
    \draw (lname) edge[-, thick] (Person);
    \draw (email) edge[-, thick] (Person);

    \draw (Person) edge[-, thick] node[above] {\it author} (wrote);
    \draw (wrote) edge[-, thick] node[above] {\it msg} (Message);

    \begin{scope}[Red]
    \node at (0.625,1.45) {$p_1$};
    \node[dot,fill,left=-2pt of University.west] (p1_1) {};
    \node[dot,fill,left=-2pt of name.west] (p1_2) {};
    \draw[->, semithick, bend angle=25] (p1_1)
      edge[bend left] node[left] {\textit{name}} (p1_2);
    \end{scope}

    \begin{scope}[Blue]
    \node at (3.75,1.25) {$p_2$};
    \node[dot,fill,left=-2pt of Person.west] (p2_1) {};
    \node[dot,fill,left=0pt of email.south west] (p2_2) {};
    \draw[->, semithick, bend angle=25] (p2_1)
      edge[bend left] node[below,sloped] {\textit{email}} (p2_2);
    \end{scope}

    \begin{scope}[Green]
    \node at (12.75,0.75) {$p_3$};
    \node[dot,fill,left=0pt of Message.north] (p3_1) {};
    \node[dot,fill,left=0pt of wrote.north] (p3_2) {};
    \node[dot,fill,left=0pt of Person.north east] (p3_3) {};
    \draw[->, semithick, bend angle=15] (p3_1)
      edge[bend right] node[above] {\textit{msg}} (p3_2);
    \draw[->, semithick, bend angle=15] (p3_2)
      edge[bend right] node[above] {\textit{author}} (p3_3);
    \end{scope}

    \begin{scope}[Orange]
    \node at (12.75,-0.75) {$p_4$};
    \node[dot,fill,left=0pt of Message.south] (p4_1) {};
    \node[dot,fill,left=0pt of wrote.south] (p4_2) {};
    \node[dot,fill,above=2pt of Person.center] (p4_3) {};
    \node[dot,fill,left=0pt of fname.south west] (p4_4) {};
    \node[dot,fill,left=0pt of lname.south] (p4_5) {};
    \draw[->, semithick, bend angle=15] (p4_1)
      edge[bend left] node[below] {\textit{msg}} (p4_2);
    \draw[->, semithick, bend angle=15] (p4_2)
      edge[bend left] node[below] {\textit{author}} (p4_3);
    \draw[->, semithick, bend angle=15] (p4_3)
      edge[bend left] node[above=-4pt, sloped] {\textit{fname}} (p4_4);
    \draw[->, semithick, bend angle=15] (p4_3)
      edge[bend left] node[below=-4pt, sloped] {\textit{lname}} (p4_5);
    \end{scope}
  \end{tikzpicture}
  \caption{Examples of patterns in a fragment of LDBC-SNB shape graph. \label{fig:patterns}}
\end{figure}
The simple pattern {\color{Red}$p_1=\textit{name}$} consists of a single attribute and
allows to retrieve the name of a university. A similarly simple pattern
{\color{Blue}$p_2=\textit{email}$} allows to access email address of a person.

The pattern {\color{Green}$p_3=\textit{msg}(\textit{author})$} allows to retrieve the person
that has authored a given message; this pattern retrieves the instances of the instances of
$\textit{Person}$ entity. We are often interested in patterns that retrieve only concrete
(attribute) values. As an alternative to the pattern $p_3$, we can use the pattern
{\color{Orange}$p_4=\textit{msg}(\textit{author}(\textit{fname}, \textit{lname}))$} that
retrieves the first and last name of the author of a message. \qed
\end{example}

Note that patterns do not need to have attribute names in all leaf nodes, but those
that do are called \emph{ground} patterns. Ground patterns are particularly
important because they allow to retrieve concrete data values (attribute values) and avoid
using identifiers of entities and relationships that are dependent on the particular
representation of a knowledge graph.

\paragraph{Syntactically valid patterns}
We only work with valid patterns that can be "embedded" in the graphical representation of
the shape graph as illustrated in Figure~\ref{fig:patterns}. For completeness in our
presentation, we formally define valid patterns as patterns rooted at an entity or relationship as follows.
A pattern $p$ \emph{is rooted at $X$ in $S$} iff
(1) $p$ is an attribute name $A$ and $X$ has attribute $A$;
(2) $p$ is a role name $R$, and either
(2.a) $X$ is a relationship with role $R$, or symmetrically
(2.b) $X$ is an entity that participates in a relationship in the role $B$;
(3) $p = B(p_1, \ldots, p_k)$, with $k \geq 1$, and either
(3.a) $X$ is a relationship with an entity $E$ participating in the role $B$, and each $p_i$
is rooted at $E$, or symmetrically
(3.b) $X$ is an entity that participates in a relationship $R$ in the role $B$, and each
$p_i$ is rooted at $R$.

\subsection{Constraints}
The constraints part of a schema $S$ are a collection of statements that we divide into a
number of categories. Again, with every statement we provide a natural language
verbalization and define the formal semantics later on.

\subsubsection{Participation Constraints}
Participation constraints specify the minimum (0 or 1) and maximum number (1 or unbounded) of
relationships instances that an entity instance participates in. They are a simple
subclass of cardinality constraints, which are not used
frequently in practice, and for which there is no general tool support. An analogous subclass of cardinality constraints allows
to specify the minimum and maximum number of attribute values of an entity or relationship
$X$ and by abusing the naming convention we put them in the same category.
\begin{description}
  \itemsep0pt
  \item $\Mandatory(X, A)$: \emph{every instance of $X$ must have an
    attribute value for $A$};
  \item $\Single(X, A)$: \emph{every instance of entity or relationship $X$ must have at most
    one attribute value of $A$};
  \item $\Mandatory(E, B ,R)$: \emph{every instance of the entity $E$ must participate
    in an instance of relationship $R$ through the role $B$};
  \item $\Single(E, B, R)$ \emph{every instance of the entity $E$ can participate in at
    most one instance of the relationship $R$ through the role $B$}.
\end{description}
\WF{2}: \emph{Every participation constraint is declared for an entity or a relationship
instance that is declared in the shape graph.} Formally, for every $\Mandatory(X, A)$ and
every $\Single(X, A)$ the shape graph contains $\Attribute(X, A)$, and for every
$\Mandatory(E, B, R)$ and every $\Single(E, B, R)$ the shape graph contains $\Role(R, B,
E)$.
\begin{example}
\label{ex:participation-constraints}
As an example, for the shape graph in Figure~\ref{fig:shape-graph} we can declare the
$\textit{date}$ attribute of the $\textit{Message}$ entity as mandatory and single with the
pair of statements $\Mandatory(\textit{Message}, \textit{date})$ and
$\Single(\textit{Message}, \textit{date})$. To declare that every
$\textit{Message}$ must have precisely one author i.e., every instance of \textit{Message} participates in the role of \textit{msg} in exactly one instance of the \textit{wrote} relationship, we can use these two statements
$\Mandatory(\textit{Message}, \textit{msg}, \textit{wrote})$ and
$\Single(\textit{Message}, \textit{msg}, \textit{wrote})$. \qed
\end{example}

\subsubsection{Key Constraints}
We introduce two kinds of key constraints, allowing to identify instances of entity or
relationship $X$. Simple key requires the uniqueness of the identifying information
specified with a collection of patterns $p_1, \ldots, p_k$. A stronger notion of identity
key requires additionally that the identifying information is always present and unique (and
so constitutes canonical identifying information). In both cases below the patterns $p_1,
\ldots, p_k$ are rooted at $X$ in the shape graph.

\begin{itemize}
  \itemsep0pt
  \item $\Key(X, [p_1, \ldots, p_k])$: \emph{no two instances of $X$ may have the same key
    values obtained with $p_1, \ldots, p_k$};
  \item $\Id(X, [p_1, \ldots, p_k])$ \emph{every instance of $X$ must have precisely one
    tuple of key values obtained with $p_1, \ldots, p_k$, and no two instances of $X$ may
    have the same key values}.
\end{itemize}
We introduce identifying keys for very specific purpose: they are used to identify instances
of entities and relationships in a manner independent of the representation of a knowledge
graph.

\WF{3}: \emph{Identity keys use ground patterns only.}

\begin{example}
\label{ex:keys}
Identity keys are naturally used to identify entities. For example, in the shape graph in
Figure~\ref{fig:shape-graph} we declare the following key: $\Id(\textit{Person},
[\textit{fname}, \textit{lname}])$.

We can use simple keys to express secondary keys, e.g., Persons may have multiple email
addresses, or even none, but no two persons should have the same email address, which can be
expressed with $\Key(\textit{Person}, [\textit{email}])$.

The use of terms in a key allows us to capture the common concept of a weak entity. For
instance, suppose that $\textit{Message}$ is a weak entity of $\textit{Person}$, which means that the identity of a message is the combination of the identity of its author and the
message number.  We can express this with the identity key $\Id(\textit{Message}, [
\textit{msg}(\textit{author}), \textit{number} ])$, or if we insist on using ground
patterns, with $\Id(\textit{Message}, [ \textit{msg}(\textit{author}(\textit{fname},
\textit{lname})) , \textit{number} ])$.

Keys can also be used on relationships, e.g., a person can be enrolled at a given university
only once in a given year, which can be expressed with the key $\Key(\textit{studies},
[\textit{uni}, \textit{student}, \textit{year}])$.
\qed
\end{example}

A standard assumption in graph theory is that graphs are simple, that is, they allow at most one edge connecting the same pair of nodes; this assumption
is transferred quite unconsciously to knowledge graphs. However, there
exist frameworks, such as property graphs and the recent RDF 1.2 \cite{W3C_RDF}, that allow multiple edges between the same pair
of nodes. For the sake of generality, we do not restrict knowledge graphs to have at most one
edge between the same pair of nodes. For instance, in the running example of the schema
(Figure~\ref{fig:shape-graph}) the same person can study at the same university multiple
times (although, at different years). Interestingly, keys allow us to impose the restriction
of singular edges if it is required for a particular application.

\begin{example}
Person following another person is a singular fact, that should not be unnecessarily
repeated, which is captured with the key $\Key(\textit{follows}, [\textit{follower},
\textit{followee}])$. \qed
\end{example}
\WF{4}: \emph{Every relationship has an identifying key.}

Note that one can use a trivial identifying key consisting of all its attributes and roles.

\subsubsection{Type Hierarchy}
It is a common and useful mechanism to organize entities into a hierarchy. For instance, in
the LDBC-SNB schema, the \textit{Message} entity has two subclasses, \textit{Post} and
\textit{Comment}. These two subclasses are disjoint: a \textit{Post} cannot be a
\textit{Comment} and vice versa. Furthermore, in this instance, the subclass relation is
total: every \textit{Message} is either a \textit{Post} or a \textit{Comment}. To support
these features we introduce the following statements.

\begin{description}
  \itemsep0pt
  \item $\Isa(E_1, E_2)$: \emph{$E_1$ is a subclass of $E_2$; in particular,  $E_1$ inherits
  all attributes, relationships, and constraints of $E_2$}.
\end{description}
\WF{5}: \emph{The type hierarchy formed with the $\Isa$ statements is acyclic}.

We clarify that multiple inheritance is not disallowed and moreover, the type hierarchy may
consist of disconnected components. To make sure that we can distinguish between entities in
the same component we introduce the following requirement.

\WF{6}: \emph{The entity at any root of the type hierarchy has an identifying key.}

Two entities from the same component of the type hierarchy can be made disjoint.
\begin{description}
  \itemsep0pt
  \item $\Disjoint(E_1, E_2)$: \emph{no instance of $E_1$ is an instance of $E_2$ and vice versa}.
\end{description}

Finally, to express total inheritance we introduce a statement that allows to specify
coverage of an entity $E$ by its descendants $E_1, \ldots, E_k$ in the type hierarchy.
\begin{description}
  \itemsep0pt
  \item $\Cover(\{E_1, \ldots, E_k\}, E)$: \emph{any instance of $E$ is an instance of at
  least one of $E_1$, \ldots, $E_k$}.
\end{description}

\begin{example}
\label{ex:type-hierarchy}
The type hierarchy among the entities \textit{Message}, \textit{Post}, and \textit{Comment}
can be expressed with the following statements:
\begin{normalfont}
\begin{align*}
  &\Isa(\textit{Post}, \textit{Message}),& &\Isa(\textit{Comment}, \textit{Message}),\\
  &\Disjoint(\textit{Post}, \textit{Comment}),& &\Cover(\{\textit{Post}, \textit{Comment}\}, \textit{Message}).\tag*{$\square$}
\end{align*}
\end{normalfont}
\end{example}

%% file: data-model.tex
\section{Data Model}
\label{sec:data-model}

Recall that the shape graph of a schema $S$ identifies the finite vocabulary $\L_S$ of
entity names $\Ent_S$, relationship names $\Rel_S$, attribute names $\Attr_S$, and role
names $\Rol_S$ used in $S$.  Also recall that, for simplicity, we work under a unique name
assumption: no attribute name and no role name are used with more than a single entity or
relationship.

We define knowledge graphs as relational structures over the union of two disjoint sets of
elements: the entity and relationship instances $\V=\V_\ent\cup\V_\rel$ and the data values
$\D$. Namely, a \emph{knowledge graph} (or simply a \emph{graph}) $G$ over $\L_S$ is a
function $\cdot^G$ that gives each element of $\L_S$ an interpretation:
\begin{align*}
  &\text{$E^G \subseteq_\fin \V_\ent $ for every $E\in\Ent_S$,}&
  &\text{$A^G \subseteq_\fin (\V_\ent \cup \V_\rel)\times\D$ for every
    $A\in\Attr_S$,}\\
  &\text{$R^G \subseteq_\fin \V_\rel$ for every $R\in\Rel_S$,}&
  &\text{$B^G \subseteq_\fin \V_\rel\times\V_\ent$ for every $B\in\Rol_S$.}
\end{align*}
Moreover, we define the sets $N_\ent = \bigcup\{E^G \mid E\in\Ent_S\} $ and $N_\rel = \bigcup \{R^G \mid
R\in\Rel_S\}$, and we require that:
\begin{enumerate}
  \itemsep0pt
  \item we do not assign attributes or roles to entity or relationship instances that do not
  belong to any entity or any relationship respectively i.e., $A^G \subseteq(N_\ent \cup
  N_\rel) \times \D$ for every $A\in\Attr_S$ and $B^G \subseteq N_\rel \times N_\ent$ for
  every $B\in\Rol_S$,
  \item role names are partial functions, i.e., if $B^G(r, e)$ and $B^G(r, e')$ then $e = e'$
  for every $B\in\Rol_S$ and $r\in R^G$,
  \item every relationship instance belongs to exactly one relationship in $\Rel_S$, i.e.,
$R_1^G \cap R_2^G = \emptyset$ for every $R_1 \neq R_2\in\Rel_S$.
\end{enumerate}
By $\G(S)$ we denote the set of all knowledge graphs over $\L_S$.

%% file: semantics.tex
\section{Semantics}
\label{sec:semantics}

Given a schema $S$ and a knowledge graph $G$ over $\L_S$, we define when $G$ satisfies $S$
by translating every statement in $S$ into a first-order logic formula.  For keys, it is
important to translate patterns into formulas that extract the identifying information. 

\subsection{Semantics of Patterns}
The translation is relatively natural but to handle correctly the direction of the role
predicates we need to know if the pattern is evaluated (rooted) at an entity or a
relationship. We illustrate this translation on a simple example below.
\begin{example}
The pattern $p_4 = \textit{msg}(\textit{author}(\textit{fname, lname}))$ from
Example~\ref{ex:patterns} is rooted at \textit{Message} and naturally translated to the
following formula:
\begin{equation*}
    \varphi_{p_4}^{\textit{Message}}(x, y_1, y_2) = \exists w, p.\ \textit{msg}(x, w) \land
    \textit{author}(w, p) \land \textit{fname}(p, y_1) \land \textit{lname}(p, y_2).
    \tag*{$\square$}
\end{equation*}

Take now the simple pattern $p_0 = \textit{student}$. Its interpretation depends on whether
it is rooted at the entity \textit{Student} or at the relationship \textit{studies}. When
rooted at \textit{Student}, we translate it to this formula:
\begin{align*}
    & \varphi_{p_0}^{\textit{Person}}(x, y) = \textit{student}(y, x)
\intertext{whereas when rooted at the relationship \textit{studies}, we translate it to this formula:}
    & \varphi_{p_0}^{\textit{studies}}(x, y) = \textit{student}(x, y). \tag*{$\square$}
\end{align*}
\end{example}

Formally, we translate a pattern $p$ of arity $k$ rooted at $X$ to a formula $\varphi_p^X(x,
y_1, \ldots, y_k)$ with the following recursive procedure:
(1) For an attribute leaf $p = A$ the formula is $\varphi_p^X(x, y) = A(x, y)$;
(2) For a role leaf $p = B$ we have either
(2.a) $X$ is an entity and the formula is $\varphi_p^X(x, y) = B(y, x)$ or, 
(2.b) $X$ is a relationship and the formula is $\varphi_p^X(x, y) = B(x, y)$;
(3) For a non-leaf pattern $p = B(p_1, \ldots, p_m)$ with $B\in\Rol_S$, we have either
(3.a) $X$ is an entity that participates in a relationship $R$ in the role $B$ and then the
formula is $\varphi_p^X(x, \bar{y}_1, \ldots, \bar{y}_m) = \exists z.\ B(z, x) \land
\varphi_{p_1}^R(z, \bar{y}_1) \land \ldots \land \varphi_{p_m}^R(z, \bar{y}_m)$ or, 
(3.b) $X$ is a relationship that participates in the role $B$ with an entity $E$ and then
the formula is $\varphi_p^X(x, \bar{y}_1, \ldots, \bar{y}_k) = \exists z.\ B(x, z) \land
\varphi_{p_1}^E(z, \bar{y}_1) \land \ldots \land \varphi_{p_m}^E(z, \bar{y}_m)$.

\subsection{Core Semantics of KG-ER}
We map each statement $\mathit{s}$ of a schema $S$ into a FOL formula $\sem{\mathit{s}}$ that captures its meaning, as presented in Figure~\ref{fig:kg-er-semantics}.
\begin{figure}[htb]
\begin{align*}
    & \sem{\Entity(X)} = \mathit{true},\qquad \sem{\Relationship(R)} = \mathit{true},\\
    & \sem{\Attribute(X, A)} = \forall x, y.\ A(x, y) \Rightarrow X(x) ,\\
    & \sem{\Role(R, B, E)} = \forall x, y.\ B(x, y) \Rightarrow R(x) \land E(y),\\
    & \sem{\Mandatory(X, A)} = \forall x.\ X(x) \Rightarrow \exists y.\ A(x, y),\\
    & \sem{\Single(X, A)} = \forall x, y, z.\ A(x, y) \land A(x, z) \Rightarrow y = z,\\
    & \sem{\Mandatory(E, B, R)} = \forall x.\ E(x) \Rightarrow \exists y.\ B(y, x),\\
    & \sem{\Single(E, B, R)} = \forall x, y, z.\ B(y, x) \land B(z, x) \Rightarrow y = z,\\
    & \sem{\Key(X, [p_1, \ldots, p_k])} = \forall x, y,
        \bar{z}_1, \ldots, \bar{z}_k.\ \psi(x, \bar{z}) \wedge \psi(y, \bar{z}) \Rightarrow x = y \text{ where } \\ 
    & \quad \psi(x, \bar{z}) = X(x) \land \textstyle\bigwedge_i \varphi_{p_i}^X(x, \bar{z}_i),\\
    & \sem{\Id(X, [p_1, \ldots, p_k])} = \sem{\Key(X, [p_1, \ldots, p_k])} \wedge \psi^{\geq 1} \wedge \psi^{\leq 1} \text{ where } \\
    & \quad \psi^{\geq 1} = \forall x.\ X(x) \Rightarrow \exists \bar{y}_1, \ldots, \bar{y}_k.\
            \textstyle\bigwedge_i \varphi_{p_i}^X(x, \bar{y}_i) \text{ and }\\
    & \quad \psi^{\leq 1} = \forall x.\ X(x) \Rightarrow \forall \bar{y}_1, \ldots, \bar{y}_k, \bar{z}_1, \ldots, \bar{z}_k.\
            \textstyle\bigwedge_i \varphi_{p_i}^X(x, \bar{y}_i) \wedge \textstyle\bigwedge_i \varphi_{p_i}^X(x, \bar{z}_i) \Rightarrow \textstyle\bigwedge_i \bar{y}_i = \bar{z}_i, \\
    & \sem{\Isa(E_1, E_2)} = \forall x.\ E_1(x) \Rightarrow E_2(x),\\
    & \sem{\Disjoint(E_1, E_2)} = \forall x.\ E_1(x) \land E_2(x) \Rightarrow \mathit{false},\\
    & \sem{\Cover(\{E_1, \ldots, E_k\}, E)} = \forall x.\ E(x) \Rightarrow E_1(x) \lor \ldots \lor E_k(x).
\end{align*}
    \caption{Core semantics of KG-ER \label{fig:kg-er-semantics}}
\end{figure}

Now, the \emph{core semantics} of schema $S$ are the knowledge graphs that satisfy all its
statements:
\begin{equation*}
    L_\circ(S) = \{G \in \G(S) \mid G \models \textstyle\bigwedge_{\mathrm{s} \in S} \sem{\mathrm{s}}\}.
\end{equation*}

It can be shown that schema reasoning in \textit{KG-ER}, namely deciding entailment among graphs,  is decidable in EXPTIME, by encoding entailment \textit{KG-ER} to the \textit{FunDL} Feature-Based Description Logics--by reifying the relationships using features as suggested in~\cite{DBLP:conf/birthday/McIntyreTW19}; this means that there can be concrete algorithms to reason with \textit{KG-ER} graphs.

%% file: disjointness.tex
\section{Disjointness and Identity}
\label{sec:disjointness}

There are three important, but related, issues when designing a conceptual data model: (Q1)
\emph{How identifiable should entities be?} (Q2) \emph{Do we have implicit disjointness of
entities without common supertypes?} and (Q3) \emph{For which entities is it possible to
indicate disjointness?}
In the data models that are used to represent knowledge graphs,
these questions are answered quite differently.  For example, in relational schemas and ER
diagrams there is usually a very strong notion of identifiability, e.g., by requiring that
all entities must have a primary key. However, in RDF and LPG schema languages there is
usually no such requirement.  Concerning implicit disjointness, this is usually assumed in
relational schemas and ER diagrams. However, in schema languages for property graphs and RDF
there is usually no such assumption.  For disjointness constraints, these are in extended ER
schema languages usually only allowed between entities that share a common supertype,
whereas in RDF schema languages there is usually no such restriction.

For Q1 we define identifiability as the ability to identify entities in terms of an entity
description that consists of the entity and a combination of associated data values. We can
distinguish three different levels of identifiability, each increasingly more demanding, but
coming with practical benefits: (1) \textbf{referenceability}, which means that for all
entity instances within a certain entity there is at least one entity description that
identifies only that entity instance, (2) \textbf{local distinguishability}, which means
that within each entity, each entity instance has exactly one such entity description, and
(3) \textbf{global distinguishability}, which means that the DBMS is able to tell for two
entity descriptions if they refer to the same entity or not.

In general global distinguishability is clearly preferred: it makes the instance of a schema
less ambiguous and makes it easier to map it faithfully and straightforwardly to a database
model such as for example a relational schema while avoiding surrogate keys or artificial
identifiers.  This is achieved for example in ORM2 by requiring that (R1) all entities have
or inherit a well-founded identifier and (R2) entities that do not have a common supertype
are assumed to be disjoint.  Note that (R1) only ensures local distinguishability, but in
combination with (R2) it also ensures global distinguishability.  After all, given two
entity descriptions it will hold that their entities have a common supertype, or not. In the
first case, the DBMS can use local distinguishability in that common supertype to decide if
the descriptions refer to the same entity, and in the second case it knows that they cannot
refer to the same entity.

For KG-ER we have decided to require R1, but not necessarily R2, since R2 may not be possible or desirable to enforce for some domains. The core semantics $L_\circ$
satisfies R1 but not R2. However, when R2 is also required, we propose an alternative
semantics $L_\bot$ with \emph{implicit disjointness}. Namely, by $\theta_S$ we denote the
conjunction of $\sem{\Disjoint(E_1, E_2)}$ for any two entities $E_1, E_2 \in \Ent_S$ that
are not directly related in the type hierarchy, i.e., do not have a common Isa-ancestor. We
can now formally define the semantics with implicit disjointness as
\begin{equation*}
    L_\bot(S) = \{G \in L(S) \mid G \models \theta_S\}.
\end{equation*}


Finally, the issue of question Q3 is answered in KG-ER by only allowing explicit
disjointness constraints between entities that have a common supertype. Under the $L_\bot$
semantics this makes sense, since disjointness constraints between entities that do not have
a common supertype would be redundant.  However, even under the $L_\circ$ semantics it makes
sense to disallow such constraints, since they might be hard to map to efficient database
schemas.

%% file: future.tex
\section{Conclusions and Future Work}
We have presented KG-ER, an expressive modeling language with the purpose of providing a unified schema language 
for knowledge graphs stored in a large variety of database systems. We have built this language through diligent 
selection of the most useful and frequently-used features while being considerate of particular features 
of existing database systems used for representing knowledge graphs. 

A formalized modeling language allows to ask and treat with the necessary rigor many interesting questions, 
some that fall into the large AI domain. Firstly, we would like to know how faithfully can we translate 
a KG-ER instance to existing database models (relational, property graphs, RDF) and their schemas, and 
conversely, can a schema of an existing database, say relational, be faithfully mapped to KG-ER? How can AI 
techniques, and LLMs in particular, offer assistance in those tasks? This leads to another interesting question: 
can KG-ER be effectively used to help to define mappings between different database models?

%% file: acknowledgements.tex
\begin{ack}
  This paper grew out of a working group at Dagstuhl Seminar 24102 on Shapes in Graph Data~\cite{dagstuhl24102}.  
  This research was supported by the European Research Council under grant
  no.~ERC-IRIS-2025-0001. The authors declare no competing interests.
 Piotr Wieczorek was supported by the National Science Centre (NCN), Poland under grant 2020/39/B/ST6/00521.
\end{ack}

%% file: appendix.tex
\section{Appendix: verbalizing KG-ER helps LLMs}
\label{sec:chatgpt}

We present a number of experiments to illustrate that using a conceptual schema can help
LLMs to tailor their answers for a number of natural and commonly used tasks. We performed
the experiments using the OpenAI's GPT-4o model on July 4th and 5th, 2025. The prompts and answers
are provided below. We point out that in our experiments multiple runs sometime returned
different answers, which is expected due to the stochastic nature of LLMs. However, in all
executions the observations were consistent: providing a conceptual schema improves the
quality of the answers.

\input{chatgpt-nlp}
\input{chatgpt-optimization}
\input{chatgpt-normalize-schema}

%% file: chatgpt-nlp.tex
\subsection{Text to query}

We begin with a task where the LLM is provided a relational schema together with a query
specification in natural language, and we ask for a corresponding SQL query.

Then, in another session, we provide a similar task where the LLM is provided with a
conceptual schema in addition to the relational schema and query specification.

\begin{chatbox}
{\bfseries Prompt:}

We are given a database whose schema is defined by the statements:

\begin{sql}
CREATE TABLE Employee(
    empl_id INT PRIMARY KEY, 
    empl_name TEXT, 
    phone TEXT, 
    address TEXT
);

CREATE TABLE Department(
    dpt_id INT PRIMARY KEY, 
    dpt_name TEXT, 
    description TEXT
);

CREATE TABLE Works_in(
    employee_id INT, 
    department_id INT, 
    FOREIGN KEY(employee_id) REFERENCES Employee(empl_id),
    FOREIGN KEY(department_id) REFERENCES Department(dpt_id),
    PRIMARY KEY(employee_id, department_id)
);
\end{sql}

Write an SQL query that counts the number of employees per department.
\tcblower

{\bfseries GPT-4o:}

To count the number of employees per department, you can use the following SQL query:

\begin{sql}
SELECT 
    d.dpt_id, 
    d.dpt_name, 
    COUNT(w.employee_id) AS num_employees
FROM 
    Department d
LEFT JOIN 
    Works_in w ON d.dpt_id = w.department_id
GROUP BY 
    d.dpt_id, d.dpt_name;
\end{sql}

\begin{gptexplanation}
### Explanation:

* **`LEFT JOIN`** ensures that departments with zero employees are still included in the result.
* **`COUNT(w.employee_id)`** counts the number of employees in each department.
* **`GROUP BY`** groups the result by department so that the count is computed per department.

This will return one row per department along with the count of employees assigned to it.
\end{gptexplanation}
\end{chatbox}

The query provided by the LLM is correct. It specifically observes that a LEFT JOIN is used
to ensure that departments without employees are included in the result.  We next issued (in
another chat) the same task with an additional conceptual schema which in particular
specifies that every department must contain at least one employee. We observe that the LLM
is able to leverage the conceptual schema and returns a simpler query which uses an internal
join rather than the left outer join:

\begin{chatbox}
{\bfseries Prompt:}

We are given a database whose schema is defined by the statements:

\begin{sql}
CREATE TABLE Employee(
    empl_id INT PRIMARY KEY, 
    empl_name TEXT, 
    phone TEXT, 
    address TEXT
);

CREATE TABLE Department(
    dpt_id INT PRIMARY KEY, 
    dpt_name TEXT, 
    description TEXT
);

CREATE TABLE Works_in(
    employee_id INT, 
    department_id INT, 
    FOREIGN KEY(employee_id) REFERENCES Employee(empl_id),
    FOREIGN KEY(department_id) REFERENCES Department(dpt_id),
    PRIMARY KEY(employee_id, department_id)
);
\end{sql}

\begin{conceptual}
The database also satisfies the following conceptual schema:

'Employee' is an entity.
'empl_id' is an attribute of the entity 'Employee'.
'empl_name' is an attribute of the entity 'Employee'.
'address' is an attribute of the entity 'Employee'.
'Department' is an entity.
'dept_id' is an attribute of the entity 'Department'.
'dept_name' is an attribute of the entity 'Department'.
'description' is an attribute of the entity 'Department'.
'Employment' is an relationship.
Every instance of the entity 'Employee' participates in one or more 
    instances of the relationship 'Employment' through the role 'works_in'.
Every instance of the entity 'Department' participates in one or more 
    instances of the relationship 'Employment' through the role 'employs'.

Write an SQL query that counts the number of employees per department.
\end{conceptual}

\tcblower

{\bfseries GPT-4o:}

\begin{markdown}
To count the number of employees per department, we need to:

1. Join the `Department` and `Works_in` tables.
2. Group the result by department.
3. Count how many employees are associated with each department.

Here is the SQL query:
\end{markdown}

\begin{sql}
SELECT 
    d.dpt_id,
    d.dpt_name,
    COUNT(w.employee_id) AS num_employees
FROM 
    Department d
JOIN 
    Works_in w ON d.dpt_id = w.department_id
GROUP BY 
    d.dpt_id, d.dpt_name;
\end{sql}

\begin{gptexplanation}
### Explanation:

* `JOIN Department d WITH Works_in w ON d.dpt_id = w.department_id`: 
    Connects departments to the employees who work in them.
* `GROUP BY d.dpt_id, d.dpt_name`: Groups the data per department.
* `COUNT(w.employee_id)`: Counts the number of employees in each department.

This query satisfies the conceptual schema's requirement 
    that every department has one or more employees.
\end{gptexplanation}
\end{chatbox}

%% file: chatgpt-optimization.tex
\subsection{Query optimization}

In this experiment, we investigate  whether LLMs can use the information in conceptual
schemas to optimize queries.

We provide the LLM with a relational schema together with an SQL query, and we ask the LLM
to optimize the query. The presented query joins two copies of \texttt{Author} to
check conditions on their birth date and country of origin. However, as
\texttt{$book_id$} is the primary key for the relationship
\texttt{Authorship}, it is possible to infer that each book can have at most one
author. Therefore, the query can be considerably simplified by removing unnecessary joins.


\begin{chatbox}

{\bfseries Prompt:}

We are given the relational schema:

\begin{sql}
CREATE TABLE Author(
    id INT PRIMARY KEY,
    first_name TEXT,
    last_name TEXT,
    birth_year INT,
    country TEXT
);

CREATE TABLE Book(
    id INT PRIMARY KEY,
    title TEXT,
    category TEXT
);

CREATE TABLE Authorship(
    book_id INT PRIMARY KEY REFERENCES Book(id),
    author_id INT REFERENCES Author(id)
);
\end{sql}

Optimize the following SQL query.
\begin{sql}
SELECT b.id, b.title
FROM 
    Book b JOIN
    Authorship w1 ON w1.book_id = b.id JOIN
    Author a1 ON a1.id = w1.author_id JOIN
    Authorship w2 ON w2.book_id = b.id JOIN
    Author a2 ON a2.id = w2.author_id
WHERE
    a1.birth_year < 2000 AND
    a2.country = 'France';
\end{sql}

\tcblower

{\bfseries GPT-4o:}

\begin{gptexplanation}
The SQL query aims to find books (b.id, b.title) that were co-authored by at least two
authors, where:

    One author (a1) was born before 2000, and

    Another author (a2) is from France.

However, the current query performs redundant joins and may produce duplicates due to
self-joins on the same tables. Let's both optimize it logically and simplify the structure.
\end{gptexplanation}

Optimized Version

\begin{sql}
SELECT DISTINCT b.id, b.title
FROM Book b
JOIN Authorship aw1 ON aw1.book_id = b.id
JOIN Author a1 ON a1.id = aw1.author_id
WHERE a1.birth_year < 2000
  AND EXISTS (
    SELECT 1
    FROM Authorship aw2
    JOIN Author a2 ON a2.id = aw2.author_id
    WHERE aw2.book_id = b.id AND a2.country = 'France'
  );
\end{sql}

\begin{gptexplanation}
Explanation of Optimization

    Avoids double join on Authorship and Author:

        Original query joins Authorship and Author twice to get a1 and a2.

        We replace the second join with an EXISTS subquery, which is generally
        faster and clearer.

    Uses DISTINCT to avoid duplicate rows caused by multiple matching authors.

    Preserves semantics:

        At least one author (a1) is born before 2000.

        At least one (possibly different) author (a2) is from France.

        Both authors must have authored the same book.

Bonus: Index Recommendations

For better performance on larger datasets, consider adding these indexes:

CREATE INDEX idx_author_birth_year ON Author(birth_year);
CREATE INDEX idx_author_country ON Author(country);
CREATE INDEX idx_authorship_book_author ON Authorship(book_id, author_id);

Let me know if you'd like to further optimize for a specific SQL engine (PostgreSQL, MySQL,
SQLite, etc.), or want to extend this to more complex logic (e.g., exact number of authors,
both conditions on the same author, etc.).
\end{gptexplanation}
\end{chatbox}

The optimized version of the query is correct, but the LLM does not infer that both conditions in the query can be checked for the single author of each book and keeps unnecessary joins. Repeated experiments generated many alternative query versions, yet none incorporated this cardinality constraint.

See additional examples of queries generated by GPT-4o below:

\begin{sql}
SELECT DISTINCT b.id, b.title
FROM Book b
JOIN Authorship w1 ON w1.book_id = b.id
JOIN Author a1 ON a1.id = w1.author_id AND a1.birth_year < 2000
JOIN Authorship w2 ON w2.book_id = b.id
JOIN Author a2 ON a2.id = w2.author_id AND a2.country = 'France';
\end{sql}

\begin{sql}
SELECT b.id, b.title
FROM Book b
WHERE EXISTS (
    SELECT 1
    FROM Authorship w1
    JOIN Author a1 ON a1.id = w1.author_id
    WHERE
        w1.book_id = b.id AND
        a1.birth_year < 2000
)
AND EXISTS (
    SELECT 1
    FROM Authorship w2
    JOIN Author a2 ON a2.id = w2.author_id
    WHERE
        w2.book_id = b.id AND
        a2.country = 'France'
);
\end{sql}

Next, in a new chat, we provide the LLM, in addition to the schema and query, with the
verbalization of the conceptual schema containing explicit cardinality constraints.

\begin{chatbox}
{\bfseries Prompt:}
We are given the relational schema:

\begin{sql}
CREATE TABLE Author(
    id INT PRIMARY KEY,
    first_name TEXT,
    last_name TEXT,
    birth_year INT,
    country TEXT
);

CREATE TABLE Book(
    id int PRIMARY KEY,
    title TEXT,
    category TEXT
);

CREATE TABLE Authorship(
    book_id INT PRIMARY KEY REFERENCES Book(id),
    author_id INT REFERENCES Author(id)
);
\end{sql}

The database satisfies also the following conceptual schema:
\begin{conceptual}
'Book' is an entity.
'id' is an attribute of the entity 'Book'.
'title' is an attribute of the entity 'Book'.
'category' is an attribute of the entity 'Book'.
'year' is an attribute of the entity 'Book'.
'Author' is an entity.
'id' is an attribute of the entity 'Author'.
'first_name' is an attribute of the entity 'Author'.
'last_name' is an attribute of the entity 'Author'.
'birth_year' is an attribute of the entity 'Author'.
'country' is an attribute of the entity 'Author'.
'Authorship' is an relationship.
Every instance of the entity 'Book' participates in exactly one instance
    of the relationship 'Authorship' through the role 'is_written_by'.
Every instance of the entity 'Author' participates in zero or more instances
    of the relationship 'Authorship' through the role 'wrote'.
\end{conceptual}

Optimize the following SQL query using the knowledge from the conceptual schema.

\begin{sql}
SELECT b.id, b.title
FROM Book b JOIN
    Authorship w1 ON w1.book_id = b.id JOIN
    Author a1 ON a1.id = w1.author_id JOIN
    Authorship w2 ON w2.book_id = b.id JOIN
    Author a2 ON a2.id = w2.author_id
WHERE
    a1.birth_year < 2000 AND
    a2.country = 'France';
\end{sql}

\tcblower

{\bfseries GPT-4o:}

We can optimize the query significantly by leveraging the conceptual schema constraint:

    Every instance of the entity 'Book' participates in exactly one instance
    of the relationship 'Authorship'.

\begin{gptexplanation}
Implication:

This means:

    Each Book has only one Authorship tuple.

    Therefore, each Book has only one Author.

Current Query Problem:

The current query joins Authorship and Author twice, assuming multiple authors per book,
which contradicts the conceptual schema. It's trying to find books where:

    a1.birth_year < 2000

    and a2.country = 'France' - on the same book

But since there's only one author per book, this condition is equivalent to asking:

    Find all books whose single author has birth_year < 2000 and country = 'France'.
\end{gptexplanation}

Optimized Query:

\begin{sql}
SELECT b.id, b.title
FROM Book b
JOIN Authorship w ON w.book_id = b.id
JOIN Author a ON a.id = w.author_id
WHERE
    a.birth_year < 2000 AND
    a.country = 'France';
\end{sql}
\begin{gptexplanation}
Benefits:

    Fewer joins: goes from 5-table join to 3-table join.

    Semantically equivalent, since each book has exactly one author.

Let me know if you want this logic applied in a query optimizer explanation or rewritten
in relational algebra.
\end{gptexplanation}

\end{chatbox}

%% file: chatgpt-normalize-schema.tex
\subsection{Schema normalization}

In this subsection, we would like to check whether LLMs can use the information in
conceptual schemas to normalize a relational schema originally provided in a denormalized
form (a single ``universal'' table joining all relationships from a star-shaped schema).  We
first present the LLM with the original schema only, without the conceptual schema.

\begin{chatbox}
{\bfseries Prompt:}
We are given a database whose schema is defined by the statements:

\begin{sql}
CREATE TABLE Employment(
    employee_id INT,
    department_id INT,
    employee_name TEXT,
    department_name TEXT,
    department_desc TEXT,
    manager INT,
    PRIMARY KEY(employee_id, department_id)
);
\end{sql}

Give a normalized schema for this database.

\tcblower

{\bfseries GPT-4o:}

\begin{sql}
CREATE TABLE Employee(
    employee_id INT PRIMARY KEY,
    employee_name TEXT
);

CREATE TABLE Department(
    department_id INT PRIMARY KEY,
    department_name TEXT,
    department_desc TEXT,
    manager INT -- could also be a foreign
                -- key to Employee if manager is an employee
);

CREATE TABLE Employment(
    employee_id INT,
    department_id INT,
    PRIMARY KEY(employee_id, department_id),
    FOREIGN KEY (employee_id) REFERENCES Employee(employee_id),
    FOREIGN KEY (department_id) REFERENCES Department(department_id)
);
\end{sql}
\end{chatbox}

We removed irrelevant summaries and explanations from the answer returned by the LLM. The relational schema provided by the LLM
is correct if we expect {\ttfamily manager} to be the unique manager of a department, but would not
fit some other cases.
The unicity and referential integrity constraints on the input (denormalized) relational schema cannot
clarify the relationship between and its manager. 
We next issued (in another chat) the same task with an additional conceptual schema which in
particular specifies that we aim to model a relationship between an employee and its possibly multiple managers.

\begin{chatbox}
{\bfseries Prompt:}

We are given a database whose schema is defined by the statements:

\begin{sql}
CREATE TABLE Employment(
    employee_id INT,
    department_id INT,
    employee_name TEXT,
    department_name TEXT,
    department_desc TEXT,
    manager INT,
    PRIMARY KEY(employee_id, department_id)
);
\end{sql}

\begin{conceptual}
The database also satisfies the following conceptual schema:

'Employee' is an entity.
'employee_id' is an attribute of the entity 'Employee'.
'employee_name' is an attribute of the entity 'Employee'.
'Department' is an entity.
'department_id' is an attribute of the entity 'Department'.
'department_name' is an attribute of the entity 'Department'.
'description' is an attribute of the entity 'Department'.
'Employment' is an relationship.
'Management' is an relationship.
Every instance of the entity 'Employee' participates in one or more 
    instances of the relationship 'Employment' through the role 'works_in'.
Every instance of the entity 'Department' participates in one or more 
    instances of the relationship 'Employment' through the role 'employs'.
Every instance of the entity 'Employee' participates in one or more 
    instances of the relationship 'Management' through the role 'managed_by'.
Every instance of the entity 'Employee' participates in one or more 
    instances of the relationship 'Management' through the role 'manages'.

Give a normalized schema for this database.
\end{conceptual}

\tcblower

{\bfseries GPT-4o:}

\begin{sql}
-- Employee Entity
CREATE TABLE Employee (
    employee_id INT PRIMARY KEY,
    employee_name TEXT
);

-- Department Entity
CREATE TABLE Department (
    department_id INT PRIMARY KEY,
    department_name TEXT,
    department_desc TEXT
);

-- Employment Relationship (works_in)
CREATE TABLE Employment (
    employee_id INT,
    department_id INT,
    PRIMARY KEY (employee_id, department_id),
    FOREIGN KEY (employee_id) REFERENCES Employee(employee_id),
    FOREIGN KEY (department_id) REFERENCES Department(department_id)
);

-- Management Relationship (manages)
CREATE TABLE Management (
    employee_id INT,     -- Employee being managed
    department_id INT,   -- Context of department
    manager_id INT,      -- Manager employee
    PRIMARY KEY (employee_id, department_id),
    FOREIGN KEY (employee_id, department_id) 
        REFERENCES Employment(employee_id, department_id),
    FOREIGN KEY (manager_id) REFERENCES Employee(employee_id)
);
\end{sql}

\begin{gptexplanation}
###  Notes:

* We **split** `Employment` and `Management` into separate tables 
    to reflect their roles as relationships.
* `Management` is modeled with a foreign key to `Employment` 
    to ensure the context of the department is preserved.
* Each `employee_id` can be managed by a `manager_id` **within a department**.

\end{gptexplanation}
\end{chatbox}

Again, we removed irrelevant summaries and explanations from the answer returned by the LLM. The LLM is able to leverage the conceptual schema and returns a normalized schema that allows an employee to have multiple managers.
We observe that the LLM chose to add a department attribute to the management relationship, an information that we had not explicitly represented in our conceptual schema.

%% file: appendix-translation.tex
\section{Appendix: translations from KG-ER to other schema formalisms}
\label{sec:kger-to-graph-schemas-translations}

In this section, we show how we can express the KG-ER conceptual schema of our running example (inspired by LDBC-SNB) into other schema formalisms.

\subsection{Our running example}
We first recap and complete our running example. We start from the shape graph from Figure~\ref{fig:shape-graph} (recalled below), then add the constraints in the paper (except the type hierarchy that we omit from this discussion for simplicity), and finally add or strengthen a few constraints to obtain a well-formed schema:

\begin{figure}[htb]
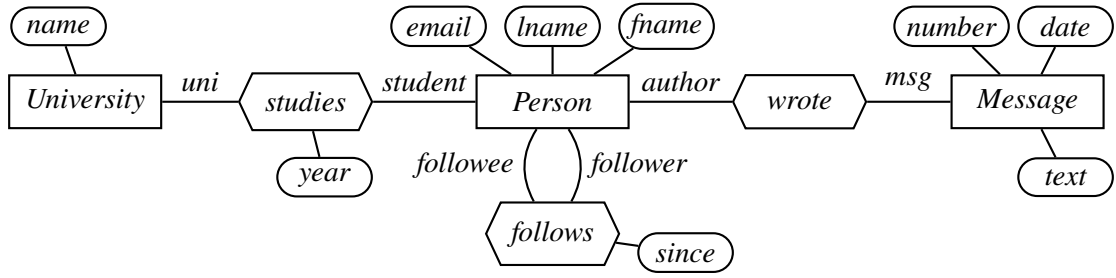

  \small
  \centering
  \usebox{\shapegraph} 
  \caption{Shape graph inspired by the LDBC-SNB schema. \label{fig:shape-graph-exact-copy}}
\end{figure}

The constraints in our example are the following:
\paragraph{Participation:} We simply collect the participation constraints from our running example, and for simplicity require that every attribute but \textit{Person(email)} is single-valued:
\begin{itemize}  \setlength\itemsep{-0.1em}
    \item $\Mandatory(\textit{Message}, \textit{date})$
    \item $\Single(\textit{Message}, \textit{date})$ and, more generally $\Single(\textit{X}, \textit{A})$ for every $X,A$ except the pair $(X=\textit{Person}, A=\textit{email})$ since we want to allow  persons to have multiple emails but want the other attributes to be single-valued. 
\end{itemize}
In addition we have:
\begin{itemize}  \setlength\itemsep{-0.1em}
  \item $\Mandatory(\textit{Message}, \textit{msg},\textit{wrote})$
  \item $\Single(\textit{Message}, \textit{msg}, \textit{wrote})$.
\end{itemize}

\paragraph{Keys:}
We not only collect the participation constraints from our running example, but also complete and strengthen keys in order to satisfy \WF{4} and \WF{6} (every relationship -- resp. "root" entity -- has an identifying key):
\begin{itemize}  \setlength\itemsep{-0.1em}

\item $\Id(\textit{University},
[\textit{name}])$

\item $\Id(\textit{Person},
[\textit{fname}, \textit{lname}])$

\item $\Key(\textit{Person}, 
[\textit{email}])$

\item $\Id(\textit{Message}, 
[ \textit{msg}(\textit{author}(\textit{fname},\textit{lname})),\, \textit{number} ])$

\item $\Id(\textit{studies},
[\textit{uni(name)},\, \textit{student(fname, lname)},\, \textit{year}])$

\item $\Id(\textit{follows}, 
[\textit{follower(fname, lname)},\, \textit{followee(fname, lname)}])$

\item $\Id(\textit{wrote},
[\textit{author(fname,lname)},\,\textit{msg(number)}])$
\end{itemize}

When translating this KG-ER schema to other schema formalisms we will for simplicity consider that every attribute has a textual data type.

\input{kger-to-relational}

\input{kger-to-shacl}
\input{kger-to-shex}
\input{kger-to-pgschema}

%% file: kger-to-relational.tex
\subsection{KG-ER to relational schemas}
In this section we present a translation of our running example KG-ER schema into a relational schema, expressed using SQL.

The translation is similar in essence to the translation from an Entity-Relationship conceptual schema to an SQL relational schema. 
Every entity type is mapped to a distinct relation. In our example, this is the case for \textit{University}, \textit{Person}, \textit{Message}.
In general, every relationship type is also mapped to a distinct relation, but some relationships can be absorbed into the participating entities depending on the constraints: \textit{studies} and \textit{follows} are expressed using a relation, but \textit{wrote} is absorbed into the \textit{Message} relation.
Single-valued attributes such as \textit{Person(fname)} are inserted into the relation which represents their parent entity or relationship, whereas a distinct relation is created for each set-valued attribute such as \textit{Person(email)}, containing the attribute and the identifying key of its parent attribute.
The constraint $\Key(\textit{Person}, 
[\textit{email}])$ which prevents two persons from having any common email is expressed in our relational schema by the unicity constraint on \textit{Emails(email)}. On our example, all KG-ER constraints translate directly into simple SQL integrity constraints. But triggers could have been required to express more complex KG-ER constraints, for instance if the patterns involved had connected distant entities through a long path.

\begin{sqlbreakable}
CREATE TABLE University(
    name text,
    PRIMARY KEY (name)
);

CREATE TABLE Person(
    fname text,
    lname text, 
    PRIMARY KEY (fname, lname)
);

-- Relation that records the set-valued attribute email:
CREATE TABLE Emails(
    email text, 
    fname text, 
    lname text,
    FOREIGN KEY (fname, lname) REFERENCES Person(fname, lname),
    PRIMARY KEY (email, fname, lname),
    UNIQUE (email)
);

/* 
Relation that records the `Message` entity 
together with its identifying relationship `works`:
*/
CREATE TABLE Message(
    author_fname text, 
    author_lname text, 
    number text, 
    date text NOT NULL, 
    msg_text text,
    FOREIGN KEY (author_fname, author_lname) REFERENCES Person(fname, lname),
    PRIMARY KEY (author_fname, author_lname, number)
);

CREATE TABLE Studies(
    university_name text,
    student_fname text,
    student_lname text,
    year text,
    FOREIGN KEY (university_name) REFERENCES University(name),
    FOREIGN KEY (student_fname, student_lname) REFERENCES Person(fname, lname),
    PRIMARY KEY (university_name, student_fname, student_lname, year)
);

CREATE TABLE Follows(
    follower_fname text,
    follower_lname text,
    followee_fname text,
    followee_lname text,
    since text,
    FOREIGN KEY (follower_fname, follower_lname) REFERENCES Person(fname, lname),
    FOREIGN KEY (followee_fname, followee_lname) REFERENCES Person(fname, lname),
    PRIMARY KEY (follower_fname, follower_lname, followee_fname, followee_lname)
);
\end{sqlbreakable}

%% file: kger-to-shacl.tex
\subsection{KG-ER to SHACL}

We present here a mapping of the running example KG-ER schema into SCHACL core, so without using SPARQL. This maps entities to nodeshapes, relationships to nodeshapes (unless they are identifying relationships such as \textit{wrote}, which are mapped to properties) and attributes to properties.

The resulting SCHACL schema fails to model fully the following constraints:
\begin{itemize}  \setlength\itemsep{-0.1em}
  \item $\Id(\textit{University}, [\textit{name}])$
  \item $\Id(\textit{Person}, [\textit{fname}, \textit{lname}])$
  \item $\Key(\textit{Person},  [\textit{email}])$
  \item $\Id(\textit{Message},  [ \textit{msg}(\textit{author}(\textit{fname},\textit{lname})),\, \textit{number} ])$
  \item $\Id(\textit{studies}, [\textit{uni(name)},\, \textit{student(fname, lname)},\, \textit{year}])$
  \item $\Id(\textit{follows}, [\textit{follower(fname, lname)},\, \textit{followee(fname, lname)}])$
\end{itemize}

Our schema features two unique keys on a single attribute: $\Id(\textit{University}, [\textit{name}])$ and $\Key(\textit{Person},  [\textit{email}])$. Such keys can in our case be expressed in SHACL using \textit{maxCount} on the reversed edge, thanks to the global uniqueness of attribute names (Section~\ref{sec:shape-graphs}).
Without this global uniqueness assumption, the unique keys on a single attribute could still be expressed in core SHACL using \textit{qualifiedMaxCount} on \textit{qualifiedValueShape} over the reversed edge, thus checking for instance that each name has a single parent entity of class University (though it may then have other parent entities of other classes). 
On the other hand, it seems it may not be possible to express composite unique keys in SHACL core \cite[End of Section 7.11]{rdfBookLabra2017}, so we did not translate the other key constraints such as $\Id(\textit{Person}, [\textit{fname}, \textit{lname}])$, though we could have expressed them using SPARQL.

\begin{lstlisting}[language=SPARQL]
@prefix sh: <http://www.w3.org/ns/shacl#> .
@prefix xsd: <http://www.w3.org/2001/XMLSchema#> .
@prefix ex: <http://example.org/> .

#################################################################
# University
#################################################################
ex:UniversityShape
    a sh:NodeShape ;
    sh:targetClass ex:University ;
    sh:property [
        sh:path ex:name ;
        sh:datatype xsd:string ;
        sh:minCount 1 ;
        sh:maxCount 1 ;
    ] .

#################################################################
# Unicity of Name globally (hence of University names)
#################################################################
ex:NameTargetShape
    a sh:NodeShape ;
    sh:targetObjectsOf ex:name ;  # all nodes that are the object of :name
    sh:property [
        sh:path [ sh:inversePath ex:name ] ;
        sh:maxCount 1 ;
    ] .

#################################################################
# Person
#################################################################
ex:PersonShape
    a sh:NodeShape ;
    sh:targetClass ex:Person ;
    sh:property [
        sh:path ex:fname ;
        sh:datatype xsd:string ;
        sh:minCount 1 ;
        sh:maxCount 1 ;
    ] ;
    sh:property [
        sh:path ex:lname ;
        sh:datatype xsd:string ;
        sh:minCount 1 ;
        sh:maxCount 1 ;
    ] ;
    sh:property [
        sh:path ex:email ;
        sh:datatype xsd:string ;
        sh:minCount 0 ;    # Optional
    ] ;
    sh:property [
        sh:path ex:studies ;
        sh:node ex:StudiesRelShape ;
        sh:minCount 0 ;
    ] ;
    sh:property [
        sh:path ex:follows ;
        sh:node ex:FollowsRelShape ;
        sh:minCount 0 ;
    ] ;
    sh:property [
        sh:path ex:wrote ;
        sh:node ex:MessageShape ;
        sh:minCount 0 ;
    ] .

#################################################################
# Unicity of emails globally (hence of Person's emails) 
#################################################################
ex:EmailTargetShape
    a sh:NodeShape ;
    sh:targetObjectsOf ex:email ;  # all nodes that are the object of :email
    sh:property [
        sh:path [ sh:inversePath ex:email ] ;
        sh:maxCount 1 ;
    ] .
    
#################################################################
# Message (Weak Entity)
#################################################################
ex:MessageShape
    a sh:NodeShape ;
    sh:targetClass ex:Message ;
    sh:property [
        sh:path ex:number ;
        sh:datatype xsd:string ;
        sh:minCount 1 ;
        sh:maxCount 1 ;
    ] ;
    sh:property [
        sh:path ex:date ;
        sh:datatype xsd:string ;
        sh:minCount 1 ;
        sh:maxCount 1 ;
    ] ;
    sh:property [
        sh:path ex:text ;
        sh:datatype xsd:string ;
        sh:minCount 1 ;
        sh:maxCount 1 ;
    ] ;
    sh:property [
        sh:path ex:writtenBy ;   
            # derived from identifying relationship "wrote"
        sh:class ex:Person ;
        sh:minCount 1 ;
        sh:maxCount 1 ;
    ] .

#################################################################
# studies relationship node (with attribute "year")
#################################################################
ex:StudiesRelShape
    a sh:NodeShape ;
    sh:property [
        sh:path ex:year ;
        sh:datatype xsd:string ;
        sh:minCount 1 ;
        sh:maxCount 1 ;
    ] ;
    sh:property [
        sh:path ex:university ;
        sh:class ex:University ;
        sh:minCount 1 ;
        sh:maxCount 1 ;
    ] .

#################################################################
# follows relationship node (with attribute "since")
#################################################################
ex:FollowsRelShape
    a sh:NodeShape ;
    sh:property [
        sh:path ex:since ;
        sh:datatype xsd:string ;
        sh:minCount 1 ;
        sh:maxCount 1 ;
    ] ;
    sh:property [
        sh:path ex:followee ;
        sh:class ex:Person ;
        sh:minCount 1 ;
        sh:maxCount 1 ;
    ] .
\end{lstlisting}

%% file: kger-to-shex.tex
\subsection{KG-ER to ShEx}

We present here a mapping of the running example KG-ER schema into ShEx. The mapping is broadly speaking the same as the one for SHACL. So we again map entities to nodeshapes, relationships to nodeshapes (unless they are identifying relationships such as \textit{wrote}, which are mapped to properties) and attributes to properties.
The result again does not fully represent all constraints of the KG-ER schema, and fails to represent the same list of constraints. Also here it is as far as we are aware not straightforward to represent in general such constraints correctly in the current version of ShEx at the time of writing, although extensions are under development that will make this possible, and in particular to express the key constraints.

\begin{lstlisting}[language=SPARQL]
PREFIX ex: <http://example.org/>
PREFIX xsd: <http://www.w3.org/2001/XMLSchema#>

# -------------------
# Entity: University
# -------------------
ex:UniversityShape {
  ex:name xsd:string          # exactly 1 name
}

# -------------------
# Entity: Person
# -------------------
ex:PersonShape {
  ex:fname xsd:string ;       # exactly 1
  ex:lname xsd:string ;       # exactly 1
  ex:email xsd:string* ;      # 0..* emails
  ex:studies @ex:StudiesRelShape* ;   # 0..* studies relationships
  ex:follows @ex:FollowsRelShape* ;   # 0..* follows relationships
  ex:wrote @ex:MessageShape*      # 0..* messages written
}

# -------------------
# Entity: Message (weak entity)
# Identified relative to a Person through 'wrote'
# -------------------
ex:MessageShape {
  ex:number xsd:string ;      # partial key (relative to Person)
  ex:date   xsd:string ;
  ex:text   xsd:string
}

# -------------------
# Relationship: studies (Person <-> University, with attribute year)
# -------------------
ex:StudiesRelShape {
  ex:year xsd:string ;
  ex:university @ex:UniversityShape
}

# -------------------
# Relationship: follows (Person <-> Person, with attribute since)
# -------------------
ex:FollowsRelShape {
  ex:since xsd:string ;
  ex:followee @ex:PersonShape
}

# -------------------
# Relationship: wrote (Person <-> Message)
# -------------------
# In ShEx we just model this via the property ex:wrote in PersonShape.
# Each Message is connected back to its Person.
\end{lstlisting}

%% file: kger-to-pgschema.tex
\subsection{KG-ER to PG-Schema}

In this section we present a translation of our running example KG-ER schema into a PG-Schema graph type as presented in \cite{PG-schema}.
The translation basically maps entities to node types and (binary) relationship to edge types. 
The exceptions are the node type \lstinline{emailType} and the edge type with label \lstinline{hasEmail}, which model the multi-valued attribute \textit{email} of entity \textit{Person}. 
The KG-ER schema constraints map straightforwardly to similarly named constraints. Note that in some cases, such as for $\Mandatory(\textit{Message}, \textit{msg},\textit{wrote})$ and $\Single(\textit{Message}, \textit{msg}, \textit{wrote})$, they are combined into a single PG-Schema constraint. 

\medskip

\begin{lstlisting}[language=PGSchema, backgroundcolor=\color{gray!10}]
  CREATE GRAPH TYPE socialStudentGraphType STRICT {
     // Node types
    (universityType: University {name STRING),
    (personType: Person {lname STRING, fname STRING}),
    (emailType: EMail {address STRING}),
    (messageType: Message {number STRING, date STRING, text STRING}),

    // Edge types
    (:personType)-[:hasEMail]->(:emailType),    
    (:personType)-[:studies {year: STRING}]->(:universityType),
    (:personType)-[:follows {since: STRING}]->(:personType),
    (:personType)-[:wrote]->(:messageType),
    
    // Node identifiers
    FOR (u:universityType) 
      IDENTIFIER u.name,
    FOR (p:personType) 
      IDENTIFIER p.fname, p.lname,
    FOR (m:messageType) 
      IDENTIFIER p.fname, p.lname, m.number WITHIN (p) -[:wrote]-> (m),

    // Edge identifiers
    FOR ()-[s:studies]->() 
      IDENTIFIER u, p, s.year WITHIN (p)-[s:studies]->(u),
    FOR ()-[f:follows]->() 
      IDENTIFIER p1.fname, p1.lname, p2.fname, p2.lname 
      WITHIN (p1)-[f]->(p2),
    FOR ()-[w:wrote]->() 
      IDENTIFIER p.fname, p.lname, m.number WITHIN (p)-[w]->(m),

    // Key and participation constraints
    FOR (m:messageType) 
      MANDATORY SINGLETON w WITHIN ()-[w:wrote]->(m),        
    FOR (p:personType)
      EXCLUSIVE e WITHIN (p)-[:hasEMail]->(e),
  }
\end{lstlisting}